\documentclass[aps,prb,twocolumn,showpacs,superscriptaddress]{revtex4}
\usepackage{graphicx}

\begin{document}

%Title of paper
\title{Ballistic magneto-thermal transport in a Heisenberg spin chain at low temperatures}
% repeat the \author .. \affiliation  etc. as needed
% \email, \thanks, \homepage, \altaffiliation all apply to the current
% author.
     \author{Lifa~Zhang}
     \affiliation{Department of Physics and Center for Computational Science and Engineering,
     National University of Singapore, Singapore 117542, Republic of Singapore }
     \author{Jian-Sheng~Wang}
     \affiliation{Department of Physics and Center for Computational Science and Engineering,
     National University of Singapore, Singapore 117542, Republic of Singapore }
     \author{Baowen~Li}
    \altaffiliation{Electronic address: phylibw@nus.edu.sg}
    \affiliation{NUS Graduate School for Integrative Sciences and Engineering,
      Singapore 117597, Republic of Singapore}
       \affiliation{Department of Physics and Center for Computational Science and Engineering,
     National University of Singapore, Singapore 117542, Republic of Singapore }
\date{19 Sep 2008}
\begin{abstract}
{We study ballistic thermal transport in Heisenberg spin chain with
nearest-neighbor ferromagnetic interactions at low temperatures.
Explicit expressions for transmission coefficients are derived for
thermal transport in a periodic spin chain of arbitrary junction
length by a spin-wave model. Our analytical results agree very well
with the ones from nonequilibrium Green's function (NEGF) method.
Our study shows that the transmission coefficient oscillates with
the frequency of thermal wave. Moreover, the thermal transmission
shows strong dependence on the intra-chain coupling, length of the
spin chain, and the external magnetic field. The results demonstrate
the possibility of manipulating spin wave propagation and
magneto-thermal conductance in the spin chain junction by adjusting
its intra-chain coupling and/or the external magnetic field.}
\end{abstract}

\maketitle

\section{Introduction}

Thermal transport properties of low-dimensional systems have gained
much attention recently\cite{review1}. Intensive studies in low
dimensional systems have made some important progress not only in
understanding the underlying physical mechanism but also in
controlling of heat. In particular, several conceptual thermal
devices have been proposed such as thermal rectifiers/diodes
\cite{rectifiers}, thermal transistors \cite{transistor}, thermal
logical gates\cite{logicgate}, thermal memory\cite{memory}, and some
molecular level thermal machines \cite{segal2006,marathe2007}. Much
work has also been done to the quantum transport in
nanostructures\cite{wangjs2008}.

Low-dimensional systems, especially one-dimensional (1D) materials,
offer the possibility to study quantum effects that are masked in
three-dimensional systems. In recent years, many interesting
experiments on thermal transport in 1D spin chains
\cite{kordonis2006,sologubenko2007,sologubenko2008,hess2007} are
performed, where the 1D-spin-chain compound materials give us nice
physical realizations of 1D toy l model systems. From these
experiments, it is possible to control heat transport in spin
systems by a magnetic field.  There are also  theoretical studies
about thermal transport in 1D spin chains, some of which show
anomalous transport due to the integrability\cite{castella1995,
saito1996,zotos1997,naef1998}, such as anisotropic Heisenberg
$S=1/2$ model, the t-V model and the $XY$ spin chain. The properties
of energy transport through the chains differ for different
anisotropy of the intra-chain spin interactions. In all spin
systems, the mean free path of itinerant spin excitations increases
as temperature decreases. Therefore in the low temperature limit,
the thermal transport in spin chain can be regarded as ballistic.
\par
The phononic transmission coefficients in quasi-one-dimensional
atomic models can be calculated by transfer matrix method
\cite{tong1999,macia2000,cao2005,antonyuk2005}. However, if there
are evanescent modes with large $|\lambda|$ ($\lambda$ is an
eigenvalue of the transfer matrix), the evaluation of the transfer
matrix can be numerically rather unstable, particularly when the
system size becomes large. Alternatively, nonequilibrium Green's
function (NEGF) is an efficient method to calculate the transmission
coefficient. Unfortunately, both of these two methods can not give
an analytical expression easily.

In this paper we give an explicit analytical expression of
transmission coefficient through a spin-wave model from a wave
scattering picture. Here we study magneto-thermal transport in an
isotropic Heisenberg spin junction coupled to two semi-infinite spin
chains in equilibrium at different temperatures.  By
Holstein-Primakoff transformation we map the spin operators to
spinless boson operators, and consider only the harmonic terms of
Hamiltonian in the low temperature limit, which is discussed in
Sec.~\ref{secmod}. The analytical solution from the spin-wave model
is shown in Sec.~\ref{secana}, we get the explicit formula for
transmission coefficient.  In Sec.~\ref{secnegf}, we introduce the
nonequilibrium Green's function method, and use it to study thermal
transport. The results and discussion  are given in
Sec.~\ref{secres}. A short summary is presented in
Sec.~\ref{seccon}.

\section{Model}
\label{secmod} The Heisenberg spin chain consists of three parts:
two semi-infinite leads and  an arbitrary junction region (see
Fig.~\ref{figmodel}). The two leads are in equilibrium at different
temperature $T_L $ and $ T_R$. We apply different external magnetic
field to the three parts along $z$ direction. So the Hamiltonian of
this system is given by
\begin{equation}\label{}
\hat H =  - \sum\limits_{i} {J_i( \hat S_i  \cdot \hat S_{i + 1}) }
- \sum\limits_i {h_i S_i^z },
\end{equation}
here $J_i$ is the interaction between spin site $i$ and $i+1$,
$h_i$ is the magnetic field applied to spin site $i$. Using
Holstein-Primakoff transformation \cite{hptransform}
\begin{equation}\label{}
S^+ = \sqrt {2S \!-\! a^ +  a}\;a;\ S^- = a^ +  \sqrt {2S \!-\! a^ +
a};\ S^z=S - a^ + a,
\end{equation}  it is easy to map spin operators to spinless
boson operators $a^ +,a $. In the low temperature limit, $ \langle
a^ + a \rangle\ll2S$, we obtain the following Hamiltonian by
neglecting the terms containing products of four or more operators,
\begin{equation}\label{}
H = E_0+ \sum\limits_{i,j} {a_i^ +  K_{ij} a_j },
\end{equation}
where $ E_0=- \sum\limits_{i} {(J_i S^2  + h_i S)} $, and $K_{ii} =
(J_{i - 1}  + J_i )S + h_i,\;K_{i,i + 1}  = K_{i + 1,i} = - J_i S$,
here $S$, which can be any integer and half-odd-integer,  is the
maximum value of spin. We choose $S=1$, without loss of generality.
\begin{figure}[h]
\includegraphics[width=3.4in,angle=0]{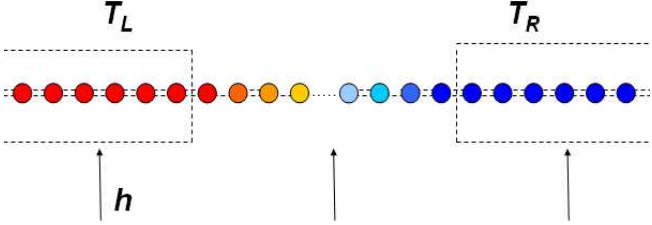}%
\label{figmodel} \caption{ \label{figmodel}(color online) The system
is an infinite Heisenberg spin chain, which consists of two
semi-infinite leads with an arbitrary junction region. The two leads
are in equilibrium at different temperature $T_L $ and $ T_R$. We
can apply different magnetic fields to the three parts. }
\end{figure}

\section{Analytical Solution from the spin-wave model}
\label{secana} Because only harmonic terms are contained in the
Hamiltonian, we can assume a spin wave solution transmitting from
the left lead to the right lead through the junction region. We
consider the two leads as uniform spin chains of intra-chain spin
coupling $J_L$  in a magnetic field $h_L$. The junction has an
alternating coupling, $J_1$ and $J_2$ in a field $h$. The unit cell
of the junction part contains two spin sites. We assume an incident
wave as $\lambda _1^j e^{ - i\omega t} $. When it arrives at the
center part, partially will be reflected and partially will be
transmitted. The reflected wave is $r\lambda _1^{ - j} e^{ - i\omega
t} $ and the transmission wave can be written as
\begin{eqnarray}
 \varphi _{j = 2m + 1}  = u\lambda _2^{j - 1} e^{ - i\omega t},  \\
 \varphi _{j = 2m}  = v\lambda _2^{j - 1} e^{ - i\omega t}.
 \end{eqnarray}
The coefficients $r$, $u$, $v$  are obtained from the continuity
condition at the interface.
 The transmission wave will be reflected and transmitted by the
right end of the junction, the amplitude of reflection wave are
$ur_{u1}$ and $vr_{v1}$ for the odd site and even one,
respectively. The transmission wave from the right end is
\begin{equation}
\varphi _j  = u\lambda _2^L t_1 \lambda _1^j e^{ - i\omega t},
\end{equation}
where $L = N - 1$, $N$ is the number of sites in the junction
part. The reflection wave will be reflected ($r_{u2}\times100\%$
percent) at the left end again, and then reflected again.
Finally the total wave function transmitted from the junction is
a superposition of multiple reflections and transmissions,
\begin{equation}\label{}
\varphi _j  = u\lambda _2^L (1 + r_{u1} r_{u2} \lambda _2^{2L}  +
(r_{u1} r_{u2} \lambda _2^{2L} )^2  +  \cdots )t_1 \lambda _1^j
e^{ - i\omega t}.
\end{equation}
\par
From time dependent Schr\"odinger equation
\begin{equation}\label{}
i\frac{\partial }{{\partial t}}\Psi  = H\Psi, \quad \quad \Psi  =
\left( {\varphi _j } \right),
\end{equation}
here we set $\hbar=1$ for simplicity, we can get the dispersion
relations for the leads $\lambda _1  = e^{iq_1 }$ and for the
junction $\lambda _2  = e^{iq_2 }$ as
\begin{eqnarray}
 \label{disper1}  \omega  - (2J_L + h_L ) =  - J_L\Bigl(\frac{1}{{\lambda _1 }} + \lambda _1 \Bigr);\quad  \\
 \label{disper2}  \bigl(\omega  - (J_1  + J_2  + h)\bigr)^2  = J_1^2  + J_2^2  + J_1 J_2 \Bigl(\frac{1}{{\lambda_2^2 }} + \lambda_2^2 \Bigr).
 \end{eqnarray}
Which root of the equations should we use?  By adding a small
imaginary part to $\omega$, that is, replacing it by $\omega + i\eta
$, then none of the eigenvalues $\lambda$ will have modulus exactly
1. Considering $\eta$ as a small perturbation, we find for the
traveling waves \cite{velev2004}
\begin{equation}\label{}
|\lambda|=1-\eta \frac{a}{v},\qquad\eta\rightarrow 0^+.
\end{equation}
That is, the forward moving waves with group velocity $ v > 0$
have $|\lambda| < 1$. In the formulas below, we take the root with
$|\lambda|<1$.
The energy band for our model ($J_1<J_2$) is
\begin{equation}\label{eband}
(\,[h,h+2J_1 )]\, \cup \,[h+2J_2,h+2(J_1+J_2 )]\,) \cap [h_L
,\,h_L+4J_L].
\end{equation}
Finally, we obtain the transmission coefficient (for $N$ odd) as
\begin{equation}\label{Tran}
\tilde T(\omega ) = \left| {\frac{{ut_1 \lambda _2^L }}{{1 -
r_{u1} r_{u2} \lambda _2^{2L} }}} \right|^2.
\end{equation}
Here,
\begin{eqnarray}
 \alpha = - \frac{{\omega  - (J_1  + J_2  + h)}}{{J_1 \lambda _2  + J_2 /\lambda _2 }} =  - \frac{{J_1 /\lambda _2  + J_2 \lambda _2 }}{{\omega  - (J_1  + J_2  + h)}}, \\
 u = \frac{{J_L(1 - \lambda _1^2 )}}{{J_L - J_2  - \lambda _1 J_L + \alpha J_2 /\lambda _2 }},\;v = u\alpha ,\\
 r_{u1} = - \frac{{\omega  - (J_L + J_2  + h) + \alpha J_2 /\lambda _2  + J_L\lambda _1 }}{{\omega  - (J_L + J_2  + h) + J_2 \lambda _2 /\alpha  + J_L\lambda _1 }},\\
 r_{u2} =  - \frac{{\omega  - (J_L + J_1  + h) + J_1 /(\alpha \lambda _2 ) + J_L\lambda _1 }}{{\omega  - (J_L + J_1  + h) + J_1 \lambda _2 \alpha  + J_L\lambda _1 }},\\
 r_{v1} = r_{u1} /\alpha ,\quad r_{v2}  = r_{u2} \alpha,\\
 t_1    = \lambda _1 (1 + r_{u1} ),\quad t_2  = \lambda _1 (1 + r_{u2} ).
\end{eqnarray}

If the number of the sites is even, that is, the length of the
chain is odd, the transmission can be written as
\begin{eqnarray}\label{Tran-even}
 \tilde T(\omega ) &=& \left| {\frac{{u\lambda _2^L t'_1 }}{{1 - r'_{u1} r_{u2} \lambda _2^{2L} }}} \right|^2 , \\
 r'_{u1}  &=& \;\alpha ^2 r_{u2} ,\quad t'_1  = \alpha t_2 .
\end{eqnarray}
If $J_1=J_2$, all the formulae are reduced to those of the uniform
spin chain. Although we only discuss the period-two spin chain, it
is easy to derive similar formulae of transmission coefficient for
any other arbitrary periodic junction by this method.
\par
For ballistic transport, the thermal current can be  written as a
Landauer-type expression:
\begin{equation}\label{eqflux}
\left\langle I \right\rangle = \frac{1}{{2\pi }}\int_{0}^\infty
{\omega \;\bigl[f_L (\omega ) - f_R (\omega )\bigr]\tilde T(\omega
)} d\omega.
\end{equation}
The conductance is
\begin{equation}\label{}
\sigma = \frac{1}{{2\pi }}\int_{0}^\infty {d\omega\;\omega\,
\tilde T(\omega )\frac{\partial f(\omega)}{\partial T}}.
\end{equation}
\begin{figure}[ht]
\includegraphics[width=3.4in,angle=0]{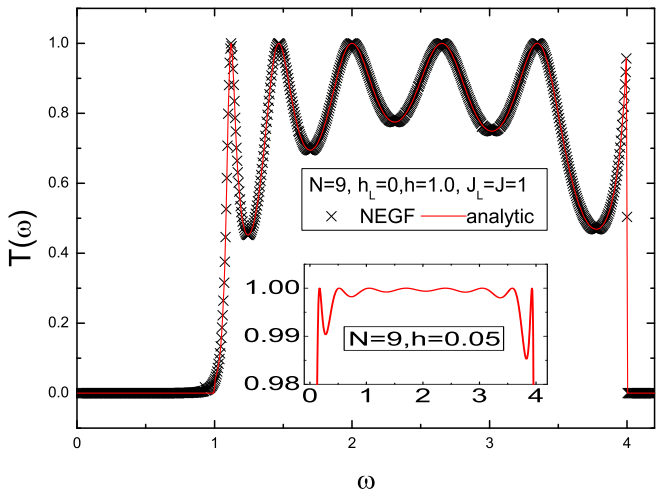}%
\label{uniform} \caption{\label{uniform}(color online) The
transmission coefficient of the uniform spin chain with coupling
$J=1$. We apply a magnetic field to the junction. The transmission
$\tilde T(\omega )$ shifts along the frequency axis with magnetic
field and oscillates with frequency in the range of $[1,4]$. The
number of sites in the junction part is $N=9$, with magnetic field
$h=1$. The scattered crosses and solid line correspond to the
results from NEGF and the spin-wave model, respectively. The inset
shows the case with a weak magnetic field, $h=0.05$. The numbers of
peaks is $N-1$.  }
\end{figure}
\section{Nonequilibrium Green's Function Method }
\label{secnegf} From the discussion in Sec.~\ref{secmod}, we can
write the Hamiltonian, by neglecting the ground state energy $E_0$,
as follows
\begin{equation}\label{}
H = \sum\limits{H_\alpha  }  + \Bigl(\sum\limits_{lm}( {a_l^{L + }
V_{lm}^{LC} a_m^C  + a_m^{C + } V_{ml}^{CL} a_l^L } ) + {\rm h.c.}
\Bigr),
\end{equation}
where $H_\alpha = \sum\limits_{lm} {a_l^{\alpha  + } K_{lm}^\alpha
a_m^\alpha},\alpha  = L,C,R$, here `$L,C,R$' denote left lead,
center part and right lead, respectively.
 The Hamiltonian matrix  of
the full linear system is
\begin{equation}\label{} H= \left( {\begin{array}{*{20}c}
   {K_L } & {V_{LC} } & 0  \\
   {V_{CL} } & {K_C } & {V_{CR} }  \\
   0 & {V_{RC} } & {K_R }  \\
\end{array}} \right).
\end{equation}
We use nonequilibrium Green's function method \cite{wangjs2008} to
study the thermal transport in the spin chain. First we define the
retarded Green's function as
\begin{equation}\label{}
G^r (t,t') =  - i\theta (t - t') \langle [a(t),a^ +  (t')]\rangle.
\end{equation}
In nonequilibrium steady states, the Green's function is
time-translationally invariant and so it depends only on the
difference in time. The Fourier transform of $G^r (t-t')=G^r
(t,t')$ is defined as
\begin{equation}\label{}
G^r[\omega]=\int_{-\infty}^{+\infty}G^r (t)e^{i\omega t}dt .
\end{equation}
We also need the advanced Green's function
\begin{equation}\label{}
 G^a (t,t') = i\theta (t' - t) \langle [a(t),a^ +  (t')] \rangle,
\end{equation}
the `greater than' Green's function
\begin{equation}\label{}
 G^ >  (t,t') =  - i \langle a(t)a^ +  (t')\rangle,
\end{equation}
and the `less than' Green's function
\begin{equation}\label{}
G^ <  (t,t') =  - i \langle a^ +  (t')a(t)\rangle.
\end{equation}
Without interaction, the free Green's functions for three parts in
equilibrium can be written as:
\begin{equation}\label{}
\begin{array}{l}
 \bigl((\omega  + i\eta ) - K_\alpha \bigr)g_\alpha ^r (\omega ) = I,\quad
\alpha  = L,C,R, \\
 g_\alpha ^a (\omega ) =g_\alpha^r(\omega)^\dagger.
  \end{array}
\end{equation}
And there is an additional equation relating $g^r$ and $g^<$:
\begin{equation}\label{}
 g^ <  (\omega ) = f(\omega )[g^r (\omega ) - g^a (\omega )],
\end{equation}
where $ f(\omega ) =  \langle a^ +  a \rangle  = [e^{\omega /T} -
1]^{-1} $ is the Bose-Einstein distribution function at temperature
$T$; we have set the Boltzmann constant $k_B=1$.
\begin{figure}[t]
\includegraphics[width=3.4in,angle=0]{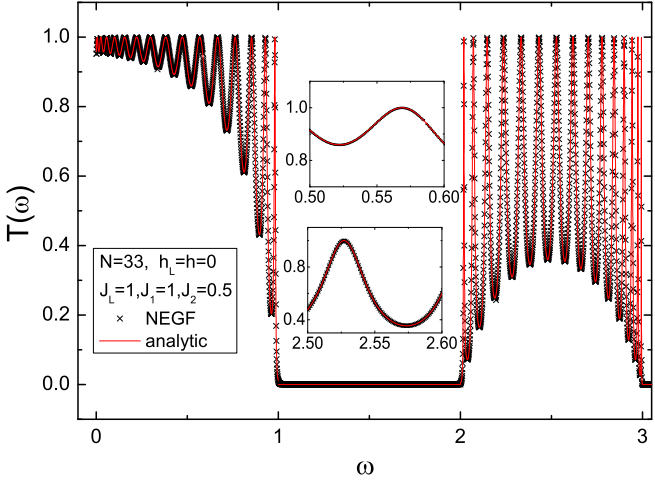}%
\label{Fig33h0} \caption{\label{Fig33h0} (color online) The
transmission coefficient of the junction part with periodic coupling
$J_1=1$ and $J_2=0.5$. There is no magnetic field applied to the
spin chain. The transmission oscillates in the band of $[0,1]$ and
$[2,3]$, which is consistent with Eq.(\ref{eband}). Here, the number
of sites in the junction part is $N=33$. The crosses and solid line
curve correspond to the results from NEGF and spin-wave model,
respectively. The inset shows the fine curves in the range of
$\omega\in[0.5,0.6]$ and $[2.5,2.6]$. The  results from two methods
are consistent with each other. }
\end{figure}
\par
The energy current flow from the left lead to the central region
is
\begin{equation}\label{}
I_L  =  -  \langle \dot H_L \rangle,
\end{equation}
which can be expressed by Green's function as
\begin{equation}\label{}
I_L = \mathop {\lim }\limits_{t' \to t^ -  }
{\rm{Tr}}\bigl\{V_{CL} K_L G_{LC}^ >  (t,t') - G_{CL}^ > (t,t')K_L
V_{LC} \big\}.
\end{equation}
In frequency domain, the current expression can be written as
\begin{equation}\label{}
I_L =  - \int_{ - \infty }^\infty \!\!\! {{\rm{Tr}}\bigl\{G_{CL}^
>  (\omega )K_L V_{LC}  - V_{CL} K_L G_{LC}^ > (\omega )\bigr\}\frac{d\omega}{2\pi}
}.
\end{equation}
Because of the following relations
\begin{eqnarray}
  K_\alpha  g_\alpha ^{ > , < }  = g_\alpha ^{ > , < } K_\alpha   = \omega g_\alpha ^{ > , < } , \\
  G_{CC}  = g_C  + g_C \Sigma G_{CC} ,\\
 \Sigma  = \Sigma _L  + \Sigma _R ,\;\Sigma _\alpha   = V_{C\alpha } g_\alpha  V_{\alpha C},\\
   G_{CL}  = G_{CC} V_{CL} g_L ,\;G_{LC}  = g_L V_{LC} G_{CC} ,
\end{eqnarray}
the current $\left\langle I \right\rangle  =
\;\frac{1}{2}(\left\langle I_L \right\rangle- \left\langle I_R
\right\rangle ) $ can be reduced to Landauer-type expression
Eq.~(\ref{eqflux}), where the transmission coefficient is
\begin{equation}\label{}
\tilde T(\omega ) = {\rm{Tr}} \bigl\{ G_{CC}^r \Gamma _R G_{CC}^a
\Gamma _L\bigr\}.
\end{equation}
 The $\Gamma _\alpha $ functions are given by $\Gamma _\alpha
= i(\Sigma _\alpha ^r  - \Sigma _\alpha ^a ).$ \\
For the 1D spin junction coupled to two semi-infinite leads, which
are uniform spin chains of intra-chain interaction $J_L$  in a
magnetic field $h_L$, the transmission coefficient can be written as
\begin{equation}\label{}
\tilde T(\omega )
=4J_L^4\bigl({\rm{Im}}(g_{00}^r)\bigr)^2\left|G_{N-1,0}\right|^2,
\end{equation}
where $ g_{00}^r=-\lambda_1/J_L$, $\lambda_1$ is given by
Eq.~(\ref{disper1}), Green's function $G_{CC}^r$ is abbreviated as
$G$ and
\begin{equation}\label{}
 G_{N - 1,0}  = (\omega  - K_C  - \Sigma )_{N - 1,0}^{ -
1}  = \frac{{( - 1)^{N - 1} \mathop \prod \limits_i J_i }}{{\det
(\omega  - K_C  - \Sigma )}}.
\end{equation}
For $N=3$, we can get $G_{2,0}  = \frac{{J_1 J_2 }}{{abc - aJ_2^2 -
cJ_1^2 }}$, here $a=\omega-J_L-J_1-h-J_L^2g_0^r$,
$b=\omega-J_1-J_2-h$, $c=\omega-J_L-J_2-h-J_L^2g_0^r$. For general
$N$, it is difficult to get an explicit formula.

\begin{figure}[t]
\includegraphics[width=3.4in,angle=0]{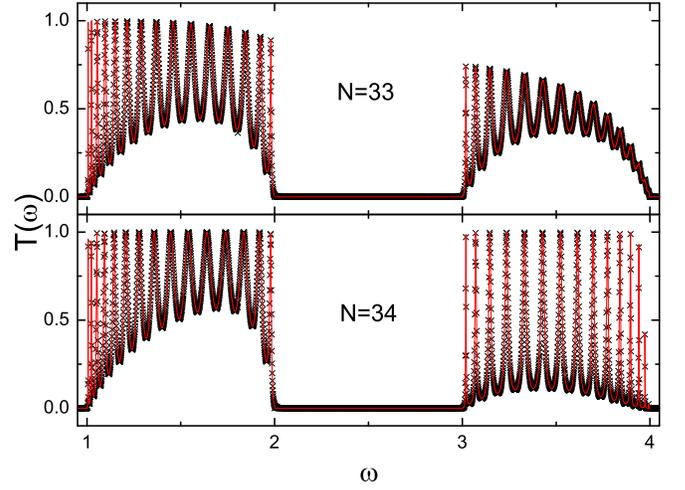}%
\label{tran-h1} \caption{\label{tran-h1}(color online) The
transmission coefficient of the periodic intra-chain coupling
junction in an external magnetic field $h=1$. Here, the coupling of
leads $J_L=1$, and $J_1=1,J_2=0.5$ for the junction. The upper and
lower panel are for $N=33$ and $N=34$, respectively. The results of
the spin-wave method (solid line curve) (for $N=33$ and $N=34$, we
use Eq.(\ref{Tran}) and  Eq.(\ref{Tran-even}), respectively) is
identical with the numerical results from NEGF (the scattered
crosses) }
\end{figure}
\section{Results and Discussions}
\label{secres}
 In our calculations, we take, $k_B=1,\;\hbar=1$. The unit of coupling
$J$ is  1\rm{meV}, then the unit of magnetic field is 17.5\rm{T};
the unit of temperature is 11.6\,K; and the unit of conductance is
$3.86 \times 10^{ - 2} {\rm{nW/K}}$.

If the whole system is uniform, i.e. the magnetic fields applied to
the three parts -- two leads and the junction -- are the same, the
transmission coefficient $\tilde T(\omega )$ is always 1 in the
whole domain $[h,h+4J]$. However, if the magnetic fields in three
parts are different, e.g., $h>0$ in the junction  and $h_L=0$ in the
two leads, the transmission coefficient oscillates with frequency
$\omega$ in the domain $[h,4J]$, where the whole system has the same
intra-chain coupling $J$. If the magnetic field is weak, the
oscillation region is very near to 1, and the number of peaks
($\tilde T(\omega )=1$) is equal to $N-1$. With the strengthening of
magnetic field in the junction, the transmission $\tilde T(\omega )$
shifts along the axis of frequency $\omega$ and the oscillation
range extends to the domain $[0,1]$, some peaks will be cut off
because the shift of curve. The numerical results come from NEGF are
exactly the same with analytical solution from the spin-wave model,
which is shown in Fig.~\ref{uniform}. All the phenomena are still
the same when the size of the spin chain is very large.  However,
for small size spin chain, the transmission at the forbidden band is
not zero because of quantum tunneling effect, which can be given
from both of the two methods and the results are exactly consistent.
The shift and oscillation of transmission coefficient is because of
interference of the spin waves transmission through the junction.

\begin{figure}[t]
\includegraphics[width=3.4in,angle=0]{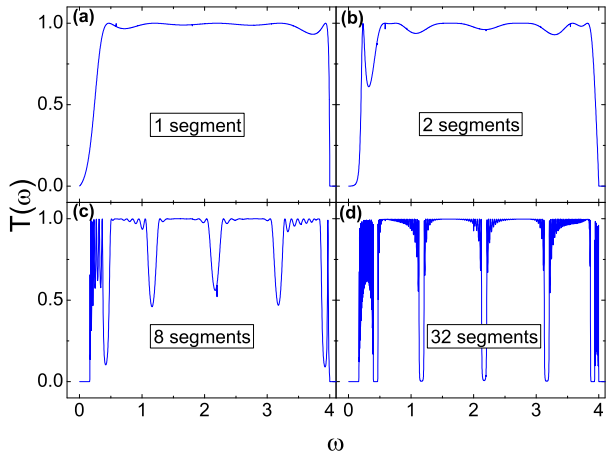}%
\label{difseg} \caption{\label{difseg}(color online) The
transmission coefficient for junctions connected in series. Here,
every segment has 5 sites and all segments in a magnetic field of
$h=0.2$ are connected by a spin with no magnetic field. (a), (b),
(c) and (d) correspond to 1,2,8 and 32 segments, respectively. For
32 or more segments, there are obvious 5 gaps (forbidden bands ),
i.e., 6 transmitted bands,which is due to the period of magnetic
field along the chain, here the period is equal to 6 (5 sites with
magnetic field plus 1 spin with no magnetic field). }
\end{figure}

\par
If the intra-chain spin coupling of junction is different from that
of leads, the transmission coefficient oscillates also with the
frequency. In Fig.~\ref{Fig33h0}, we show the transmission
coefficient for the junction with periodic coupling: $J_1=1$,
$J_2=0.5$ and  the leads with $J=1$, oscillates with frequency in
the energy band $[0,1]$ and $[2,3]$ which is consistent to
Eq.(\ref{eband}). From this figure, it can be further concluded that
our analytical results from the spin-wave model is exact. The
numerical results from NEGF method is consistent with this
analytical approach.
\par

\begin{figure}[t]
\includegraphics[width=3.4in,angle=0]{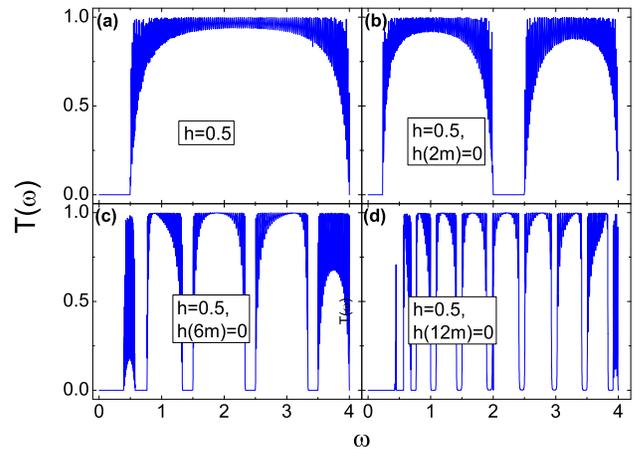}%
\label{difper} \caption{\label{difper}(color online) The
transmission coefficient for a fixed junction in different periods
of magnetic field. The total number of sites is $N=191$. The
magnetic field is applied to the junction as $h(p*m)=0,m=1,2,3...,
{\rm{others}}=0.5$,  $p$ is the period of magnetic field along the
junction. (a), (b), (c) and (d) correspond to uniform, $p=2$, $p=6$
and $p=12$ magnetic field, respectively. It is shown that there are
$p-1$ gaps for different cases.}
\end{figure}

When an external magnetic field $h=1$ is applied to the junction
part, the transmission coefficient shifts to $[1,2]\bigcup[3,4]$,
and the oscillation shape changes, which is shown in
Fig.\ref{tran-h1}. The shapes of oscillation are different for
odd-site and even-site junction. For the even-site junction, the
spin chain is symmetric along the chain direction, while it is
asymmetric for odd-site case. Therefore, the reflection in two ends
of the junction for asymmetric chain are different, which causes the
difference of the transmission compared with symmetric case.

\par
From the above results, we know that the periodicity of the junction
can give rise to gaps in the transmission. Can we merely apply
magnetic field periodically to the junction to induce gaps in the
transmission, while the whole system have the same coupling $J=1$,
therefore, we can choose the frequency to transmit from the
junction? It is possible, because that the transmission oscillates
with frequency when the junction is applied a magnetic field. If we
connect many junctions in series, then the range of oscillation
extends and may give gaps in transmission. In Fig.~\ref{difseg}, at
first we let the heat transfer through a 5-site junction in a
magnetic field $h=0.2$, the transmission oscillates a little near 1;
if we connect two junctions together by a site without magnetic
field, that is, $N=11, h(i=1\sim5;7\sim11)=0.2, h(6)=0$, the
oscillation will be extended. The gaps are shown evidently when 32
or more segments are connected in series by the sites without
magnetic field. For a fixed size junction, if we apply magnetic
field periodically, there are $p-1$ gaps ( $p$ is the period of
magnetic field along the junction) in the transmission, which are
shown in Fig.~\ref{difper}. Therefore, we can choose the frequencies
to transmit through the junction by adjusting the periodicity of the
intra-chain coupling or the magnetic field.
\begin{figure}[ht]
\includegraphics[width=3.4in,angle=0]{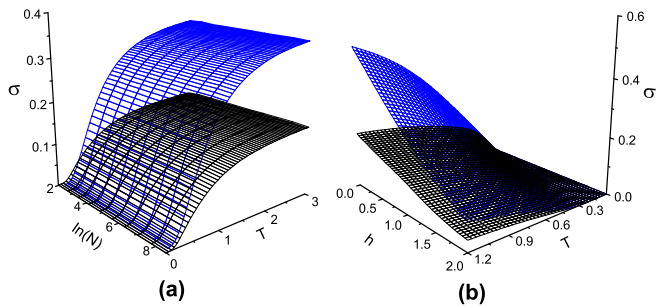}%
\label{figcond} \caption{\label{figcond}(color online)(a) The
thermal conductance, $\sigma$, versus system size, $N$, and
temperature $T$. $\sigma$ increases with temperature to a constant
whereas it keeps invariant with system size, where magnetic field is
$h=0.5$. (b) The thermal conductance $\sigma$ versus magnetic field,
$h$ and temperature. It shows that the conductance decreases to zero
with strengthening magnetic field at different temperature. The
upper blue surface and lower black surface correspond to uniform
($J=1$) and periodic ($J_1=1,J_2=0.5$) spin chains, respectively.}
\end{figure}

\par
We can calculate the thermal conductance from transmission
coefficient. In Fig.~\ref{figcond}, we show the thermal conductance
versus system size, temperature and magnetic field for uniform and
periodic spin chain. It is shown that the conductance is independent
of the system size because of ballistic transport while the
oscillation of transmission coefficient changes with the system
size. And the conductance increases to a constant with the increase
of temperature, which is consistent with all other ballistic cases.
From Fig.~\ref{figcond}(b), we see that the conductance decreases to
zero as the intensity of magnetic field is increased. These results
indicate that the transmission coefficient will shift along the axis
of frequency, which cuts off the contribution of low frequency to
the thermal current, therefore the thermal current decreases.
Because the thermal wave of low frequency contributes most to the
heat flux, the thermal conductance decreases quickly with the
increase of the intensity of magnetic field. Although the
transmission has a big difference for uniform and periodic chains,
the thermal conductance has the similar behavior merely with a
difference of magnitude.

\section{Conclusion}
\label{seccon} In this paper, we have studied the ballistic
magneto-thermal transport in a Heisenberg spin chain at low
temperatures. We have obtained explicitly an analytical expression
for the transmission coefficient through a spin-wave model from a
wave scattering picture. The analytical results have been verified
by the nonequilibrium Green's function method.  We have found that
the transmission coefficient oscillates with the frequency because
of interference of the transmission waves; furthermore, the thermal
transmission coefficient shows strong dependence  not only on the
intra-chain coupling and length of the spin chain but also on the
external magnetic field. There are gaps in the transmission, the
number of gaps is equal to $p-1$, where $p$ is the period of
intra-chain coupling or the external transverse magnetic field
applied to the junction, i.e., the number of transmitted bands is
equal to the value of periods. Therefore, it is easy to choose
special frequencies to transmit through the spin chain junction by
adjusting its intra-chain coupling or the external magnetic field
and the heat current in the junction can be  switched off with the
magnetic field strengthening. The thermal conductance of Heisenberg
spin chain at low temperature tends to a constant with the
temperature increasing, decreases to zero with intensity of the
magnetic field, while it has no dependence on the system size.

Our analytical spin-wave model solution can be applied to the
ballistic magneto-thermal transport in an arbitrary periodic spin
chain. The properties of the magneto-thermal transport found in this
paper provide the possibility to manipulate magneto-thermal
conductance and the propagation of spin waves in the Heisenberg spin
chain,  which may have potential applications in thermal control and
designing of filter and waveguide for spin waves.

\subsection*{Acknowledgements}
We thank Yonghong Yan for fruitful discussions. The work is
supported by the grant R-144-000-203-112.  J.-S Wang acknowledge
support from a faculty research grant R-144-000-173-112/101 of
NUS.


\begin{thebibliography}{99}
\bibitem{review1}
D. G. Cahill, W. K. Ford, K. E. Goodson, G. D. Mahan, A. Majumdar,
H. J. Maris, R. Merlin, and S. R. Phillpot, J. Appl. Phys. {\bf 93},
793 (2003).
\bibitem{rectifiers}Theoretical papers on lattices please refer to:
M. Terraneo, M. Peyrard, and G. Casati, Phys. Rev. Lett.{\bf 88},
094302 (2002); B. Li, L. Wang, and G. Casati, Phys. Rev. Lett. {\bf
93}, 184301 (2004); B Li, J.-H Lan, L Wang, Phys. Rev. Lett. {\bf
95}, 104302 (2005); D. Segal and A. Nitzan, Phys. Rev. Lett. {\bf
94}, 034301 (2005); B. Hu, L. Yang, and Y. Zhang, Phys. Rev. Lett.
{\bf 97}, 124302 (2006); N. Yang, N.- B Li, L Wang, and B Li, Phys.
Rev. B {\bf 76}, 020301 (2007). Theoretical work on nanosture
rectifiers please refer to: G Wu and B. Li, Phys Rev. B {\bf
76}£¬085424 (2007), G. Wu and B. Li, J. Phys.: Condens. Matter {\bf
20}, 175211 (2008);M. Hu, P. Keblinski and B. Li, Appl. Phys. Lett
{\bf 92}, 211908 (2008). Experimental one please refer to: C. W.
Chang, D. Okawa, A. Majumdar, and A. Zettl, Science {\bf 314}, 1121
(2006); R. Scheibner, M. K\"onig, D. Reuter, A.D. Wieck, H. Buhmann,
and L.W. Molenkamp, New J. Phys. { \bf 10} 083016 (2008).
\bibitem{transistor}
B. Li, L. Wang, and G. Casati, Appl. Phys. Lett. {\bf 88}, 143501
(2006); W. C. Lo, L. Wang, and B. Li, J. Phys. Soc. Jpn. {\bf 77},
054402 (2008).
\bibitem{logicgate}
L. Wang and B. Li, Phys. Rev. Lett. {\bf 99}, 177208 (2007); L. Wang
and B. Li, Physics World {\bf 21}, No.3, 27-29 (2008).
\bibitem{memory}L. Wang and B. Li, arXiv0808:3311.v1.
\bibitem{segal2006}
D. Segal and A. Nitzan, Phys. Rev. E {\bf 73}, 026109 (2006).
\bibitem{marathe2007}
R. Marathe, A. M. Jayannavar, and A. Dhar, Phys. Rev. E {\bf 75},
030103 (2007).
\bibitem{wangjs2008} J.-S. Wang, J. Wang and J. T. L\"{u}, Eur. Phys. J. B {\bf 62},
381 (2008).
\bibitem{kordonis2006}
K. Kordonis, A. V. Sologubenko, T. Lorenz, S.-W. Cheong, and A.
Freimuth, Phys. Rev. Lett. {\bf 97}, 115901 (2006).
\bibitem{sologubenko2007}
A. V. Sologubenko£¬ K. Berggold, T. Lorenz, A. Rosch, E.
Shimshoni, M. D. Phillips, and M. M. Turnbull, Phys. Rev. Lett.
{\bf 98}, 107201
 (2007).
\bibitem{sologubenko2008}
 A. V. Sologubenko£¬ T. Lorenz, J. A. Mydosh, A. Rosch, K. C. Shortsleeves, and M. M. Turnbull, Phys. Rev. Lett. {\bf 100}, 137202
 (2008).
\bibitem{hess2007}
  C. Hess, Eur. Phys. J. Special Topics {\bf 151}, 73-83 (2007).
\bibitem{castella1995}
H. Castella, X. Zotos, and P. Prelov{\v{s}}ek, Phys. Rev. Lett.
{\bf 74},972 (1995).
\bibitem{saito1996}
K. Saito, S. Takesue, S. Miyashita, Phys. Rev. E. {\bf 54},2404
(1996).
\bibitem{zotos1997}
X. Zotos, F. Naef, and P. Prelov{\v{s}}ek, Phys. Rev. B. {\bf
55},11029 (1997).
\bibitem{naef1998}
 F. Naef and X. Zotos, J. Phys. :Condens. Matter {\bf 10}, L183 (1998).
\bibitem{tong1999}
 P. Tong, B. Li, B. Hu, Phys. Rev. B {\bf  59}, 8639 (1999)
 \bibitem{macia2000}
 E. Maci{\'{a}}, Phys. Rev. B {\bf 61}, 6645 (2000)
  \bibitem{cao2005}
 L. S. Cao, R. W. Peng, R. L. Zhang, X. F. Zhang, Mu Wang, X. Q. Huang, A. Hu, and S. S. Jiang, Phys. Rev. B {\bf 72}, 214301 (2005)
 \bibitem{antonyuk2005}
 V. B. Antonyuk, M. Larsson, A. G. Mal¡¯shukov, K. A. Chao, Semicond. Sci. Technol. {\bf 20}, 347 (2005)
\bibitem{hptransform}
T. Holstein and H. Primakoff, Phys. Rev. {\bf 58}, 1098 (1940).
\bibitem{velev2004}
 J. Velev, W. Butler, J. Phys.: Condens. Matter {\bf 16}, R637 (2004)
\end{thebibliography}
\end{document}